\documentstyle[aps,prl,preprint,epsf]{revtex}
\newcommand{\etal}{{\em et al.}}                

\newcommand{\ibid}{{\em ibid }}
\newcommand{\dn}[2]{d^{#1}{#2}\,}
\newcommand{\dnvec}[1]{d \vec{#1}\,}
\newcommand{\expval}[1]{\left< #1 \right>}
\newcommand{\eqref}[1]{(\ref{#1})}
\newcommand{\PSDens}{\expval{f(y,p_T)}}
\newcommand{\PSDensalt}{\expval{f(\vec{p})}}
\newcommand{\avePSDens}{\expval{f}}

\newcommand{\fdens}{f(\vec{r},\vec{p})}
\newcommand{\ffdens}{f^2(\vec{r},\vec{p})}
\newcommand{\Spectrum}{\frac{\dn{2}{N}}{2\pi\dn{}{y}\dn{}{p_T}p_T}}
\newcommand{\Spectrumalt}{\frac{E\dn{3}{N}}{\dnvec{p}}}
\newcommand{\Spectrumaltalt}{\frac{\dn{3}{N}}{\dnvec{p}}}
\newcommand{\Stozero}{S_{\vec{p}}(\vec{r'}\rightarrow 0)}
\newcommand{\Npart}{N_{\rm part}}
\begin{document}
\title{
  \begin{flushright}{\rm DOE/ER/40561-82-INT00-INT}\\[9mm]\end{flushright}
Extracting particle freeze-out phase-space densities and entropies from 
sources imaged in heavy-ion reactions}
\author{David A. Brown$^{1}$, Sergei Y. Panitkin$^{2}$ and 
George F. Bertsch$^{1}$}
\address{\em $^1$Institute for Nuclear Theory, 
University of Washington, Box 351550, Seattle, WA 98195-1550}
\address{\em $^2$Department of Physics, Kent
State University, Kent, Ohio 44242}
\date{\today}
\maketitle
%
\begin{abstract}
The  space-averaged  phase-space density  and  entropy  per  particle are  both
fundamental  observables   which  can   be  extracted  from   the  two-particle
correlation  functions measured  in heavy-ion  collisions.  
%
%
%
%
Two techniques have been proposed to extract the densities from correlation
data: either by using the radius parameters from Gaussian fits to meson 
correlations or by using source imaging, which may be applied to any like pair 
correlation.  We show that the 
imaging and Gaussian fits give the same result in the case of meson 
interferometry.  We discuss the concept of an equivalent instantaneous 
source on which both techniques rely. 
%
%
%
%
%
We  also discuss the
phase-space occupancy and entropy per particle.
Finally, we propose an improved formula for  the phase-space occupancy
that has a  more  controlled  dependence on  the
uncertainty of the experimentally measured source functions.
\end{abstract}
%
\section{Introduction}
The  phase-space  density  of   particles  produced  in  an  ultra-relativistic
heavy-ion collision is  a fundamental observable which is  accessible, at least
in part,  via two-particle correlations.   Measurements of this  observable are
interesting  because  they  may  either  provide direct  evidence  for  thermal
phase-space  distribution  of  particles  at  freeze-out or  show  evidence  of
deviations from  such~\cite{e877,Ferenc:1999ku}.  Also, if one  can measure the
average  phase-space  density,  one  can   begin  to  look  for  effect  of  an
overpopulation of  phase-space or  more exotic phenomena  such as  pion lasers,
superradiance, etc. \cite{pi_lasers}.  
%
%
A quantity closely related to the phase-space density is the entropy per 
particle,  a key thermodynamic property 
of high density matter.  Indeed, a phase transition might not take place 
adiabatically and could generate entropy.
%
%

Several methods were suggested for the measuring the phase-space
densities of the various particles in a heavy-ion  collision.  For the  case of
identical  noninteracting pions,  Bertsch~\cite{bertsch_HBT} proposed  a method
that uses  both the  radius parameters from  conventional HBT analysis  and the
pion  spectrum.   He  also  suggested  a  physically  intuitive  definition  of
phase-space density:  replace the single  particle sources in  the Koonin-Pratt
formalism~\cite{koonin_77,HBT:pra90}  with an  instantaneous source  at  a mean
freeze-out  time.  This  effective  source is  just  the effective  phase-space
density  at freeze-out.   This replacement  preserves the  particle  numbers at
large times  (after freeze-out) even  though, by construction, it  differs from
the true density during the freeze-out process.

Another approach, based on the source imaging technique introduced by Brown and
Danielewicz   \cite{HBT:bro97,HBT:bro98},    allows   one   to    extract   the
space-averaged  phase-space density  and entropy  per particle  from  {\em any}
like-pair  correlations.  This approach  uses both  the source  function imaged
from a correlation measurement and  the single particle spectrum.  We will show
that  this approach  is more  versatile:  it can  be applied  to any  like-pair
correlations  and  it reduces  to  the  HBT  result of~\cite{bertsch_HBT}  when
applied to pions.

A third approach, introduced by Siemens and Kapusta~\cite{dpratios},  
uses the ratio of total deuteron yield to total proton yield instead of the
two-particle correlation function to estimate the total 
phase-space occupancy of nucleons.  This approach is complimentary to the 
source imaging approach in the sense that both give access to the proton 
phase-space occupancy.  However, while the imaging method requires one to 
measure the correlation function and one particle spectrum for one specie 
of particle, the deuteron-proton ratio approach requires one to measure 
the spectra of two different species.  
Usually these spectra have different acceptances and efficiencies of
particle reconstruction and identification which may complicate
experimental analysis.
We will not discuss this approach in  this letter.

The outline  of the  paper is as  follows.  First,  we use the  substitution of
Ref.~\cite{bertsch_HBT}  to explain  how  the  imaged sources  can  be used  to
extract the space-averaged phase-space density, $\PSDensalt$, as discussed 
in~\cite{HBT:bro97}.  We will show that these
results   for   $\PSDensalt$   are   a   generalization  of   the   result   in
Ref.~\cite{bertsch_HBT} derived for identical mesons.  Next, we will demonstrate
that one does  not change the source function by  making this substitution.  
Within this framework, we will discuss 
the phase-space occupancy, $\avePSDens$,  and the entropy per 
particle, ${\cal S}/\Npart$.   
We will find an  expression  for  the  the  phase-space
occupancy   that is an   improvement  over   that
in~\cite{HBT:bro97} as the  formula here has a more  controlled dependence on
the uncertainty of the experimentally measured source functions.

\section{The Correlation and Source Functions}

We begin by defining a Lorentz-invariant single particle source, 
$D(r,\vec{p})$, which gives the rate for creating on-shell 
($E=\sqrt{\vec{p}^2+m^2}$) particles (of all spin projections):
\begin{equation}
	D(r,\vec{p})=\frac{E\dn{7}{N}}{\dnvec{r}\dn{}{t}\dnvec{p}}
\label{eqn:psource}
\end{equation}
In our notation,  
$r=(t,\vec{r})$ is a four-vector and $\vec{p}$ is a three vector.  
With our choice of normalization, the single-particle source transforms 
as a Lorentz scalar.  The single-particle source may be computed directly 
in an event generator such as RQMD~\cite{sorge_95}.    

We define the two-particle source function as the probability density for 
producing a particle pair separated by $\vec{r'}$ in their center of mass (CM) 
frame. Following~\cite{koonin_77,HBT:pra90,HBT:bro97,HBT:bro98},  we define 
this source  function  as the convolution of the single-particle sources:
\begin{equation}\begin{array}{rl}
	S_{\vec{p}}(\vec{r'})
	    &=\frac{\displaystyle\int dt'\int \dn{4}{R}
		D(R+r/2,\vec{p}_1)D(R-r/2,\vec{p}_2)}
		{\displaystyle\left(\int \dn{4}{R}D(R,\vec{p}_1)\right)
		\left(\int \dn{4}{R}D(R,\vec{p}_2)\right)}\\
	    &\equiv\displaystyle\int dt'\int \dn{4}{R}
		\tilde{D}(R+r/2,\vec{p}_1)\tilde{D}(R-r/2,\vec{p}_2)
\end{array}\label{eqn:thesource}\end{equation}
Here we denote the normalized single-particle sources with a tilde, 
coordinates taken in the pair CM frame with primes, and coordinates 
taken in the the lab frame without primes.  The  average particle
momentum   is   $\vec{p}=\frac{1}{2}(\vec{p}_1+\vec{p}_2)$   and   the
relative particle momentum is $\vec{q}=\frac{1}{2}(\vec{p}_1-\vec{p}_2)$.     
While all $\vec{q}$ pairs seem to enter in this 
definition~\eqref{eqn:thesource}, in practice
only low relative momentum pairs contribute to correlations and hence either to 
the imaged sources or Gaussian fits to correlations.  Typically, 
this is used to allow one to replace $\vec{p}_1$ and $\vec{p}_2$ with the 
average pair momentum $\vec{p}$ (see Ref.~\cite{HBT:pra90}).
Following convention, we will not explicitly state the $\vec{q}$ dependence  
of the  source function.

The two-particle source function may be extracted from the measured 
two-particle correlation 
function by inverting the Koonin-Pratt~\cite{koonin_77,HBT:pra90} equation:
\begin{equation}
	C_{\vec{p}}(\vec{q'})=
	\frac{\displaystyle\frac{\displaystyle d N_{\rm pair}}
	{\displaystyle d\vec{p}_1d\vec{p}_2}}
	{\displaystyle\frac{\displaystyle d N}{\displaystyle d\vec{p}_1}
	 \displaystyle\frac{\displaystyle d N}{\displaystyle d\vec{p}_2}}=
	\int\dn{3}{r'}\left|\Phi_{\vec{q'}}(\vec{r'})\right|^2 
	S_{\vec{p}}(\vec{r'})
\label{eqn:PKeqn}
\end{equation}
Here  $C_{\vec{p}}(\vec{q'})$   is  the  measured   correlation  function,  and
$\Phi_{\vec{q'}}(\vec{r'})$ is  the pair relative  wavefunction in the  pair CM
frame.  We comment that the correlation function can be measured and the source
function extracted in any frame, however relation between the two is simplest 
in the pair CM frame. 

In order to extract the source function, we first
discretize Eq.~\eqref{eqn:PKeqn} to obtain the matrix equation
$C_i=\sum_j K_{ij} S_j$.  We then proceed as
in~\cite{HBT:bro97,HBT:bro98} and find the
set of source points, $S_j$, that minimize the $\chi^2$.  Here, 
$\chi^2=\sum_i(C_i - \sum_j K_{ij} S_j)^2/\Delta^2C_i$.
The $\chi^2$ minimizing source is $S_j=\sum_i[(K^TBK)^{-1}K^TB]_{ji} (C_i-1)$ 
where $K^T$ is the transpose of the kernel matrix and $B$ is 
the inverse covariance matrix of the data
$B_{ij}=\delta_{ij}/\Delta^2C_i$.
The error on the source is the square-root of the diagonal elements of the
covariance matrix of the source, $\Delta^2S=(K^TBK)^{-1}$.
Since this procedure works on any like-pair correlation, 
we can dispense with the correlation function and work directly with the  
source function.  

\section{The Phase-Space Density}

The phase-space density at time $t$ is a Lorentz scalar; we write it as
\begin{equation}
	f(t,\vec{r},\vec{p}) = 
	\frac{1}{\Gamma_0}\frac{\dn{6}{N}}{\dnvec{r}\dnvec{p}}.
\end{equation}
Here we have  defined a   unit   volume   in   phase-space   as
$\Gamma_0=(2s+1)/(2\pi\hbar  c)^3$ and a  differential 
phase space volume  as $\Gamma_0\dnvec{r}\dnvec{p}$.  The $(2s+1)$ factor
accounts for the spin of the particle of interest.  
Assuming that the particles propagate as free particles,
the  phase-space density at  a specific
time $t_1$ can be written in terms of the phase-space density a different 
time $t_0$ as
\begin{equation}\begin{array}{lr}
	\lefteqn{f(t_1,\vec{r},\vec{p}) = 
	f(t_0,\vec{r}-\vec{v}(t_1-t_0),\vec{p})}&\\
	&+\displaystyle \frac{1}{\Gamma_0 E}\int_{t_0}^{t_1}\dn{}{t}
	D(t,\vec{r}-\vec{v}(t_1-t),\vec{p})
\end{array}\label{eqn:dandf}\end{equation}
This expression ignores  processes that modify the
particle's  momentum (such as  from a mean-field,  Coulomb forces,  etc.). 
Since we  want the  particles at  position $\vec{r}$, we  must look  backwards 
(or forwards)    in   time    to   where    it   was    at its   creation,
e.g.  $\vec{r}-\vec{v}(t_1-t)$.

A particle is said to have frozen out when it has undergone its last strong 
interaction and is now propagating to the detector.  Since each particle can 
freeze-out at a different time, the concept of a ``freeze-out phase-space 
density'' is ambiguous.
One option  to deal  with this  is to average  over the  creation times  of the
particles.  However then the question becomes how to perform this average and 
what the averaged density means.  Ref. \cite{bertsch_HBT} proposes 
a simple solution:  replace   the  particle  source $D(r,\vec{p})$
with  an effective  source $D^{\rm  eff}(r,\vec{p})$ at an instantaneous 
freeze-out time, $t_f$.  In other words,
\begin{equation}
	D(r,\vec{p})\rightarrow  
	D^{\rm  eff}(r,\vec{p})=\delta(t_{f}-t)  E  \Gamma_0
	f^{\rm eff}(\vec{r},\vec{p})
\label{eqn:instantD}
\end{equation}
The  effective freeze-out  phase-space density is related  to the  true 
single-particle source via:
\begin{equation}
	f^{\rm  eff}(\vec{r},\vec{p})=\frac{1}{\Gamma_0   E}  
	\int_{-\cal  T}^{\cal  T}
	\dn{}{t} D(t,\vec{r}-\vec{v}(t_f-t),\vec{p})
\label{eqn:effectivef}
\end{equation}
Here, ${\cal T}>t_f>-{\cal T}$ and 
$\pm{\cal T}$  is simply some large time after  (before) which the source
is turned off (on).   It is clear from~\eqref{eqn:effectivef} that 
replacement in~\eqref{eqn:instantD} gives the 
right phase-space density at  large times ($t>{\cal T}$) as it must by 
Liouville's Theorem.  Furthermore, this can be easily checked by 
examining Eq.~\eqref{eqn:dandf} for large times
$t>{\cal T}$ with $f=0$ for times earlier than ${\cal -T}$.
Finally, it is clear that this replacement does not alter the momentum space 
density that one might calculate from Eq.~\eqref{eqn:dandf}.

Not only does this prescription give a reasonable way to define the
freeze-out phase-space density, but it also is {\em entirely consistent with our
definition of the source function in Eq.~\eqref{eqn:thesource}}.  
We will illustrate this by simply inserting the effective phase-space
density into the definition of the source (Eq.~\eqref{eqn:thesource}) and
performing the required algebra.  Special care must be taken in performing the 
various time integrals as some are performed in the lab frame while others are
performed in the pair CM frame.  

We now insert Eqs.~(\ref{eqn:instantD},\ref{eqn:effectivef}) into the equation 
for the source function in 
Eq.~\eqref{eqn:thesource} and perform one of the delta function
integrals.  We obtain
\begin{equation}\begin{array}{rl}
	S_{\vec{p}}(\vec{r'})=&\displaystyle\int\dn{}{t'}\delta(t)
	\int\dn{3}{R}\int\dn{}{T}\int\dn{}{\tau}\\
	&\displaystyle\times
	\tilde{D}(T+\tau/2,\vec{R}+\vec{r}/2-\vec{v}_1(t_f-T-\tau/2),\vec{p}_1)
	\\
	&\displaystyle\times
	\tilde{D}(T-\tau/2,\vec{R}-\vec{r}/2-\vec{v}_2(t_f-T+\tau/2),\vec{p}_2).
\end{array}\label{eqn:midstep1}\end{equation}
Here $\vec{v}_i=\vec{p}_i/E_i$ are the velocities of the individual particles. 
Also, $T=(t_1+t_2)/2$ and $\tau=(t_1-t_2)$ are the average and relative time
variables from the time integrals in the effective phase-space density. 
We can simplify Eq.~\eqref{eqn:midstep1} by introducing the average velocity of 
the pair: $\vec{v}\equiv\vec{p}/\sqrt{\vec{p}^2+m^2}$ so that 
$\vec{v}_i=\vec{v}+\delta \vec{v}_i$.  With this, we make the change of 
variables $\vec{R}-\vec{v}(t_f-T)\rightarrow\vec{R}$ and remove much of the 
time dependence from the spatial arguments:
\begin{equation}\begin{array}{rl}
	S_{\vec{p}}(\vec{r'})=&\displaystyle\int\dn{}{t'}\delta(t)
	\int\dn{3}{R}\int\dn{}{T}\int\dn{}{\tau}\\
	&\displaystyle\times
	\tilde{D}(T+\tau/2,\vec{R}+(\vec{r}+\vec{v}\tau)/2-\delta\vec{v}_1(t_f-T-\tau/2),\vec{p}_1)\\
	&\displaystyle\times
	\tilde{D}(T-\tau/2,\vec{R}-(\vec{r}+\vec{v}\tau)/2-\delta\vec{v}_2(t_f-T+\tau/2),\vec{p}_2)
\end{array}\end{equation}

Our next step is to do the $t'$ integral.  This appears to  
be a straightforward delta function integral, however care must be taken due 
to the different reference frames involved.  When we perform the $t'$ 
integral, two things happen: 
$\int \dn{}{t'}\delta (t)\rightarrow 1/\gamma$ and everywhere in the 
integrand $t'\rightarrow -vr_\|'$.  Here $r_\|'$ is the component of $\vec{r'}$ 
in the direction of the boost from the lab to the pair CM frame (this boost
velocity is $\vec{v}$).

We make further progress by examining $\vec{r}+\vec{v}\tau$.  First,
define the four-vector
$s=(\tau,\vec{s})=(\tau,\vec{r}+\vec{v}\tau)$.
If we write this in the pair CM frame, we have
$s=\left(\gamma(\tau'+vs_\|'),\gamma(s_\|'+v\tau'),\vec{r'}_\perp\right)$.
So, the parallel component of $s$ is 
\begin{equation}
s_\|=\gamma(s_\|'+v\tau')=\gamma(s_\|'+v(\frac{\tau}{\gamma}-vs'_\|))
=\frac{1}{\gamma}s'_\|+v\tau
\end{equation}
But note, from the definition of $s$ and from the $t'$ integral we already have 
\begin{equation}
s_\|=r_\|+v\tau=\gamma(r_\|'+vt')+v\tau=\frac{1}{\gamma}r_\|'+v\tau.
\end{equation}
Therefore, if we identify $s_\|'=r_\|'$ and change integration variables 
from $\tau$ to $\tau'$, we find the result
\begin{equation}\begin{array}{rl}
	S_{\vec{p}}(\vec{r'})=&\displaystyle\int\dn{}{\tau'}
	\int\dn{3}{R}\int\dn{}{T}\\
	&\displaystyle\times
	\tilde{D}(T+\tau/2,\vec{R}+\vec{r}/2-\delta\vec{v}_1(t_f-T-\tau/2),\vec{p}_1)\\
	&\displaystyle\times
	\tilde{D}(T-\tau/2,\vec{R}-\vec{r}/2-\delta\vec{v}_2(t_f-T+\tau/2),\vec{p}_2).
\end{array}\label{eqn:checkbertsch}\end{equation}

In order to complete the connection between Eq.~\eqref{eqn:checkbertsch} 
and the source in Eq.~\eqref{eqn:thesource}, we must justify dropping the
$\delta\vec{v}_i$'s in Eq.~\eqref{eqn:checkbertsch}.
Since $\Delta t=(t_f-T\mp t/2)$ is on the order of the freeze-out 
duration, if $\delta\vec{v}_i\Delta t$ is smaller than the characteristic
length scale of the single-particle source, we can drop the 
$\delta\vec{v}_i$'s.
Writing $\delta\vec{v}_i$ in terms of the relative momentum $\vec{q}$, we find 
\begin{equation}
\delta\vec{v}_i=\vec{v}-\vec{v}_i
=\pm \frac{1}{E}(\vec{q}-\vec{v}\cdot\vec{q}\vec{v})+{\cal O}(\vec{q}\,^2).
\label{needaname}
\end{equation}
Thus, the $\delta\vec{v}_i$ term shifts the spatial argument of $\tilde{D}$ in 
Eq.~\eqref{needaname}
$\sim\Delta t q/\gamma E$ in the direction parallel 
to $\vec{v}$ and $\sim\Delta t q/E$ perpendicular to it.  For highly 
relativistic or massive pairs, this shift may be neglected.   
For low velocity or light pairs (such as $\pi$ pairs) the shift is 
important, especially if the freeze-out duration is large or the system size is
small.  

\subsection{Space-Averaged Phase-space Density from the Sources}

Inserting   the  particle   source    with   instantaneous    freeze-out   into
Eq.~\eqref{eqn:thesource},  taking $\vec{p}_1, \vec{p}_2\approx \vec{p}$,
taking  the limit  as  $\vec{r'}\rightarrow 0$,  and
performing the integrals over time, we find Eq.~(17) of Ref.~\cite{HBT:bro97}:
\begin{equation}
  \PSDensalt\equiv\frac{\displaystyle\int\dn{3}{r}\ffdens}
  {\displaystyle\int\dn{3}{r}\fdens}=\frac{1}{\Gamma_0}\frac{1}{m}\Spectrumalt
  \Stozero
\label{eqn:altPSdens}
\end{equation}
For the sake of brevity, we drop  both the ``eff''  tag on the density 
here and for the remainder of the paper. 

For zero impact parameter collisions, we may exploit the azimuthal symmetry and
average  over the  angle of  the particle  transverse momentum  $\theta_{p_T}$ 
and use
\begin{equation}
\begin{array}{rl}
\displaystyle\PSDens &\equiv
	\frac{\displaystyle\int\dn{}{\theta_{p_T}}\int\dn{3}{r}\ffdens}
	{\displaystyle\int\dn{}{\theta_{p_T}}\int\dn{3}{r}\fdens}\\
&=\displaystyle\frac{1}{\Gamma_0}\frac{1}{m}\Spectrum\Stozero
\end{array}
\label{eqn:PSdens}
\end{equation}
Note  that, due to  the azimuthal  symmetry the  density is  a function  of the
particle rapidity $y$ and the magnitude of the particle transverse momentum, 
$p_T$.

As a  side comment on  using Eqs.~\eqref{eqn:altPSdens} and~\eqref{eqn:PSdens},
one  can  use either  the  $\vec{r'}\rightarrow 0$  point  from  either a  full
three-dimensional   reconstruction   of    the   source   function   (such   as
in~\cite{APSmeeting}) or from the angle-averaged source function.  In practice,
it  is  usually  much  easier   to  measure  the  angle-averaged  two  particle
correlation   function   (and   hence    source   function)   then   the   full
three-dimensional  correlation because  one  can sum  over  angles to  increase
statistics.

\subsection{Gaussian Meson Sources}
\label{gaussians}

We now show that Eqs.~\eqref{eqn:PSdens} and~\eqref{eqn:altPSdens} are direct 
generalizations of  the results in~\cite{bertsch_HBT} or~\cite{Ferenc:1999ku} 
for pions.  
The correlation  function for identical non-interacting spin-$0$
bosons   can  be   written  in  terms  of  a   matrix  of  radius
parameters
\cite{wiedemann_99}:
\begin{equation}
	C(\vec{Q})= 1+\lambda e^{-Q_iQ_j[R^2]_{ij}}.
	\label{eqn:gaussiancorr}
\end{equation}
Here we have dropped the average pair momentum label and all primes on 
the momenta.  Here  also, $\vec{Q}=2\vec{q}$ (the  relative
momentum variable used in  the analysis of pion correlations) and
$\lambda$ is a fit parameter often called the chaoticity parameter.
The  matrix of  radius parameters,  $[R^2]$, is  the following  real, symmetric
matrix:
\begin{equation}
	[R^2]=\left(
	\begin{array}{ccc}
		R_o^2 & R_{os}^2 & R_{o\ell}^2 \\ 
		R_{os}^2 & R_s^2 & R_{s\ell}^2 \\ 
		R_{o\ell}^2
		& R_{s\ell}^2 & R_\ell^2
	\end{array}
	\right)
\end{equation}
in the Bertsch-Pratt parameterization.
For pions, the Koonin-Pratt equation  is a Fourier cosine transform that
may  be  inverted analytically~\cite{HBT:bro97}  to  give  the source  function
directly in terms of the correlation function:
\begin{equation}
	S(\vec{r})=\frac{1}{(2\pi)^3}\int\dn{3}{Q}
           \cos{(\vec{Q}\cdot\vec{r})}
	\left(C(\vec{Q})-1\right).
\end{equation}
Inserting Eq.~\eqref{eqn:gaussiancorr}  into this expression  yields a Gaussian
source function:
\begin{equation}
	S(\vec{r})=\frac{\lambda}{(2\sqrt{\pi})^3\sqrt{\det{[R^2]}}}
	\exp{\left(-\frac{1}{4}r_ir_j[R^2]^{-1}_{ij}\right)}
\end{equation}
Taking the $r\rightarrow 0$ limit of this source
\begin{equation}
	S(\vec{r}\rightarrow 0)=\frac{\lambda}{(2\sqrt{\pi})^3\sqrt{\det{[R^2]}}}
\end{equation}
and  inserting this  into the  equation  for the  average phase-space  density,
Eq.~\eqref{eqn:altPSdens} or~\eqref{eqn:PSdens}  yields the result 
in Ref.~\cite{bertsch_HBT}.
Actually our result is more general than those in 
Refs.~\cite{Ferenc:1999ku,bertsch_HBT} as 
those results only apply either to diagonal  $[R^2]$ or $[R^2]$ with 
$R_{o\ell}^2 \ne 0$.  
One  should note that  the $\vec{r}=0$ intercept  of the
source  function  has  units of  an  effective  volume.

\section{Phase-Space Occupancy and Entropy}

We can now estimate $\avePSDens$ 
from our calculation  of the space-averaged phase-space 
density, $\PSDens$ or $\PSDensalt$.  In~\cite{HBT:bro97},
the authors argue that $\avePSDens$ can be estimated via
\begin{equation}
	\avePSDens = \frac{\displaystyle\int\dn{3}{p}\PSDensalt^2}
	{\displaystyle\int\dn{3}{p}\PSDensalt}.
	\label{eqn:avePSden1}
\end{equation}
Given the current state of available correlation data, the uncertainty 
in the extracted sources can be greater than $50\%$ of the extracted source 
value.  Therefore, in this expression the uncertainty in the result 
is dominated by the uncertainty in the underlying extracted source.
Indeed, since the relative uncertainty in
$\PSDensalt$ is nearly 
$\delta  S_{\vec{p}}(r\rightarrow 0)/S_{\vec{p}}(r\rightarrow 0)$,  we find
$\delta \avePSDens/\avePSDens \sim 3 \delta  S_{\vec{p}}(r\rightarrow 0)/S_{\vec{p}}(r\rightarrow 0)$
for sources with a strong $\vec{p}$ dependence.
Thus, one easily finds uncertainties greater than  the  values themselves.
Clearly an alternative is needed that  has a smaller dependency in the error on
the source.  

Instead   of~\eqref{eqn:avePSden1},  we  propose   the  following   method  for
evaluating $\avePSDens$:
\begin{equation}
	\avePSDens=\displaystyle\frac{1}{\Npart}\int\dn{3}{p}
	\PSDensalt\Spectrumaltalt,
\label{eqn:avePSden2}
\end{equation}
and for azimuthally symmetric systems,
\begin{equation}
	=\displaystyle\frac{2\pi}{\Npart}
	\int\dn{}{y}\dn{}{p_T}p_TE\PSDens\Spectrum.
\label{eqn:avePSden3}\end{equation}
Using either of these expressions, the relative uncertainty goes like one 
factor of $\delta  S_{\vec{p}}(r\rightarrow 0)/S_{\vec{p}}(r\rightarrow 0)$.  
One can  see \eqref{eqn:avePSden2} (or \eqref{eqn:avePSden3}) by 
beginning  with the  definition of $\avePSDens$:
\begin{equation}\begin{array}{rl}
	\avePSDens&=\frac{\displaystyle\int\dn{3}{r}\dn{3}{p}\ffdens}
		{\displaystyle\int\dn{3}{r}\dn{3}{p}\fdens}\\
		&=\frac{\displaystyle\int\dn{3}{p}
		\left(\PSDensalt\int\dn{3}{r}\fdens\right)}
		{\displaystyle\Npart/\Gamma_0}\\
		&=\frac{\displaystyle\int\dn{3}{p}
		\left(\PSDensalt\frac{1}{\Gamma_0}
		\frac{\dn{3}{N}}{\dnvec{p}}\right)}
		{\displaystyle\Npart/\Gamma_0}
\end{array}\end{equation}
which immediately gives Eq.~\eqref{eqn:avePSden2}.

One might wonder if, through similar considerations, we may be able to 
improve on the calculation of the entropy per particle given in 
Ref.~\cite{HBT:bro97} or~\cite{dpratios}.  In short, we do not believe so.  
Consider the entropy for a gas of fermions (top) or bosons (bottom):
\begin{equation}
	{\cal  S}/\Npart= -\frac{\displaystyle\int\dn{3}{r}\dn{3}{p}
	\left[f\ln{f}\pm(1\mp f)\ln{(1\mp f)}\right]}
	{\displaystyle\int\dn{3}{r}\dn{3}{p}f}.
\end{equation}
To arrive at their expression for the entropy per particle, the authors
of~\cite{HBT:bro97} neglected the spatial dependence of 
$f(\vec{r},\vec{p})$ and found:
\begin{equation}\begin{array}{rl}
	\lefteqn{{\cal  S}/\Npart=}&\\
	&-\frac{\displaystyle\int\dn{3}{p}\left[\PSDensalt\ln{\PSDensalt}\pm
	(1\mp\PSDensalt)\ln{(1\mp\PSDensalt)}\right]}
	{\displaystyle\int\dn{3}{p}\PSDensalt}.
\end{array}\label{eqn:entropy1}\end{equation}
The authors of~\cite{dpratios} go one step farther and neglect the momentum
variation of $f(\vec{r},\vec{p})$ to obtain a result entirely dependent 
on the phase-space occupancy $\left<f\right>$.
Clearly the entropy should get larger contributions from regions of coordinate
space where the phase-space density is small however both 
Eq.~\eqref{eqn:entropy1} and the analogous result from Ref.~\cite{dpratios}
do not reflect this.   The source function's $\vec{r}$ dependence does 
give us information about the spatial dependence of the phase-space density
however it is not clear how one might use this information to obtain a better
estimate of the entropy. 

Now we present sample calculations of $\avePSDens$ and ${\cal  S}/\Npart$ from 
two different experiments.  The first calculation uses negative pion 
correlations measured from Pb+Pb collisions at 158~GeVA from 
the CERN-SPS experiment NA49 \cite{NA49,appelsthesis,schoenthesis}.  The second 
calculation uses proton correlations from the 
$^{14}$N$+^{27}$Al reaction at $75$~MeVA 
measured at the Michigan State University NSCL \cite{msupp}.  
These calculations will show the general applicability of the source 
imaging and the superiority of Eq.~\eqref{eqn:avePSden2} over 
Eq.~\eqref{eqn:avePSden1}.  

For the first calculation, we use the space-averaged
$\pi^-$ phase-space densities of Ferenc et al.~\cite{Ferenc:1999ku}, 
extracted from experiment NA49 \cite{NA49,appelsthesis,schoenthesis}.  
In this calculation we could have used an extracted source, but we can 
also get the $\vec{r'}\rightarrow 0$ source intercept from the radius 
parameters of a Gaussian fit to the pion correlations, as we saw in 
section~\ref{gaussians}.  Since this is exactly what is used 
in Ref.~\cite{Ferenc:1999ku}, 
we use their radii.  We estimate the $\pi^-$ spectrum
as the product of the $\pi^-$ rapidity and $p_T$ distributions:
\begin{equation}
	\frac{dN_{\pi^-}}{dy dp_T p_T}=0.9\frac{dn_{N^-}}{dy}
	\left[T_{\rm eff}(y)(m_\pi+T_{\rm eff}(y))\right]^{-1}
	\exp\left(-\frac{\sqrt{p_T^2+m_\pi^2}-m_\pi}{T_{\rm eff}(y)}\right)
	\label{myspectrum}
\end{equation}
Here, $dN_{h^-}/dy$ is the negative hadron rapidity distribution from 
Ref.~\cite{appelsthesis} and the factor of $0.9$ accounts for the fraction of 
negative hadrons that are actually pions.  In~\eqref{myspectrum}, the 
$p_T$ distribution is parameterized by a rapidity dependent effective 
temperature $T_{\rm eff}(y)$ and the actual values of this effective temperature
are obtained from~\cite{schoenthesis}.  Using~\eqref{eqn:avePSden2}, we find 
$\avePSDens=0.19\pm 0.06$ while using~\eqref{eqn:avePSden1}, we find 
$\avePSDens=0.14\pm 0.08$.  Both results are consistent, however the result from
\eqref{eqn:avePSden2} has a $25\%$ smaller uncertainty.  Using 
Eq.~\eqref{eqn:entropy1}, the entropy per
pion can also be estimated as ${\cal  S}/N_{\pi^-}=3.9\pm 1.8$.

The numbers extracted from the sources from the NSCL pp data in
\cite{msupp} are more dramatic.  Since the proton 
spectrum for this reaction is only available at a few angles, we follow 
Ref.~\cite{HBT:bro97} and approximate it with the thermal 
distribution:
\begin{equation}
	\frac{dN_p}{\dnvec{p}}\propto\frac{1}{z^{-1}\exp\left(p^2/2mT\right)+1}.
\end{equation}
Here the normalization constant is determined by normalizing to the number of
participants ($9$ protons), $T$ is the fitted temperature of $10.2$~MeV, 
and z is set from the requirement of maximum entropy giving $z\sim 1.1$.
Since the source is given only at three fixed pair momenta, we follow 
Ref.~\cite{HBT:bro97} and average the results to obtain the zero intercept
$S_{\rm ave} (r\rightarrow 0)=15.5\pm 2.5\times 10^{-4}$ fm$^{-3}$.
Using~\eqref{eqn:avePSden2}, we find $\avePSDens=0.27\pm 0.12$ while 
using~\eqref{eqn:avePSden1}, we find $\avePSDens=0.25\pm 0.04$ -- a factor
of 3 improvement in the uncertainty.  
For the entropy per proton,  we estimate ${\cal  S}/N_{p}=2.7\pm 0.7$.
These results differ slightly from those in~\cite{HBT:bro97} simply because 
the authors of~\cite{HBT:bro97} set the error on the source to zero.  
Using the more accurately determined intercepts from Ref.~\cite{HBT:bro98}, 
$S_{\rm ave} (r\rightarrow 0)=18.7\pm 1.1\times 10^{-4}$  fm$^{-3}$, we find
$\avePSDens=0.30\pm 0.02$ from~\eqref{eqn:avePSden2} and 
$\avePSDens=0.30\pm 0.05$ from~\eqref{eqn:avePSden1}.  For the entropy we find
${\cal  S}/N_{p}=2.45\pm 0.21$.  Note the substantial improvement 
from a better determination of the source intercept.  Nevertheless, the result
from Eq.~\eqref{eqn:avePSden2} is still a factor of 2 
improvement over~\eqref{eqn:avePSden1}.

\section{Conclusions}

The  phase-space  density is  an  important,  fundamental  observable that  can
provide insight  into the underlying dynamics  of the nuclear  reactions and it
may be extracted  from sources imaged in heavy-ion  reactions.  This extraction
relies on  an appropriate  definition of the  effective phase-space  density at
freeze-out,  since  the true  freeze-out  density  is  not a  uniquely  defined
concept.  We  have  shown that  the definition in Ref.~\cite{bertsch_HBT} is
entirely consistent  with the  source function obtained by 
imaging~\cite{HBT:bro97}.  We have also shown how the 
extraction of the space-averaged phase-space density from
imaged sources is a generalization of Bertsch's result for pion correlations.
Finally, we have provided a formula for the calculation of
the average phase-space occupancy.  This formula
is less sensitive to the uncertainties  of the  source
functions  than  others  in  the   literature.   
We believe that the source imaging method will be useful for extracting 
space-averaged phase-space densities from data of future 
relativistic heavy-ion experiments.


\section*{Acknowledgements}
\indent We gratefully acknowledge stimulating discussions with 
Drs. A.~Parre{\~n}o, N.~Xu, F.~Wang, T.~Papenbrock, K.~Hagino, S.~Reddy, 
L.~Kaplan and J.~Cramer.  This research is supported by the U.S. Department of
Energy grants  DOE-ER-40561 and DE-FG02-89ER40531. 
%

%
\end{document}